\pdfoutput=1

\documentclass[11pt]{article}

\usepackage[preprint]{acl}

\usepackage{times}
\usepackage{latexsym}

\usepackage[T1]{fontenc}

\usepackage[utf8]{inputenc}

\usepackage{microtype}

\usepackage{inconsolata}

\usepackage{graphicx}

\usepackage{hanging}

\usepackage{pifont}

\title{Theories of ``Sexuality'' in Natural Language Processing Bias Research}

\author{Jacob T. Hobbs \\
  Department of Computer Science \\
  Department of Women, Gender, and Sexuality \\
  University of Virginia\\
  \texttt{fdu5ff@virginia.edu} \\}

\begin{document}
\maketitle
\begin{abstract}



In recent years, significant advancements in the field of Natural Language Processing (NLP) have positioned commercialized language models as wide-reaching, highly useful tools. In tandem, there has been an explosion of multidisciplinary research examining how NLP tasks reflect, perpetuate, and amplify social biases such as gender and racial bias. A significant gap in this scholarship is a detailed analysis of how queer sexualities are encoded and (mis)represented by both NLP systems and practitioners. Following previous work in the field of AI fairness, we document how sexuality is defined and operationalized via a survey and analysis of 55 articles that quantify sexuality-based NLP bias. We find that sexuality is not clearly defined in a majority of the literature surveyed, indicating a reliance on assumed or normative conceptions of sexual/romantic practices and identities. Further, we find that methods for extracting biased outputs from NLP technologies often conflate gender and sexual identities, leading to monolithic conceptions of queerness and thus improper quantifications of bias. With the goal of improving sexuality-based NLP bias analyses, we conclude with recommendations that encourage more thorough engagement with both queer communities and interdisciplinary literature.

\end{abstract}

\section{Introduction}


In a landmark essay that bridges the gap between Queer Theory and practical work in Artificial Intelligence, Nishant Shah questions attempts to mesh queer life with machine binarisms, observing the following:

\begin{quote}
\textit{"The rhetoric of AI development as necessarily improving the human condition, but particularly removing the “unwanted” or “undesirable” structures of contamination and corruption, inevitably frames queerness as a site of detection, management, containment, and punishment, thus falling in a long legacy of technological refusal to recognise it as a legitimate subculture of lifestyle, and measuring it always as an aberration."}

\hfill -- \citet{shah_i_2023}
\end{quote}
This framing exposes how queerness is often misrepresented, made invisible, or pathologized by AI. Bias against queer people in Natural Language Processing (NLP), then, is not merely a technical flaw or glitch but a reflection and reinforcement of systemic prejudice. With the growing prominence of language model technology, anti-queer bias in NLP systems can inflict even greater harms against an already vulnerable population. This can manifest as both "allocational harms," where queer people are discriminated against or treated differently compared to other groups, or "representational harms," where queerness is misrepresented or erased within NLP systems \citep{gallegos_bias_2024}.

In an effort to mitigate these harms, sexuality is increasingly becoming a common demographic axis for which to measure bias in NLP systems. Building on seminal work uncovering algorithmic racial and gender bias \citep{bolukbasi_man_2016, caliskan_semantics_2017, manzini_black_2019, sun_mitigating_2019}, a handful of authors have begun to research how queer sexualities are processed in various NLP tasks such as language generation \citep{sheng_woman_2019, dhingra_queer_2023, felkner_gpt_2024}, machine translation \citep{stewart_whose_2024}, sentiment analysis \citep{ungless_potential_2023}, hate speech/toxicity detection \citep{davani_hate_2023}, and even heteronormativity detection \citep{vasquez_heterocorpus_2022}. Scholars in the field of gender and sexuality studies have noted, however, that a stable definition of queerness is slippery at best, owing to the social construction of sexuality and its entanglement with constricting systems of power \citep{foucault_history_1978, butler_gender_1990}. Butler in particular emphasizes how language itself plays an important role in constructing identity categories in the first place. Therefore, we are primarily interested in how technical researchers articulate, codify, and thus reinforce various theories of sexuality, centering our analysis on NLP bias research. 


Specifically, we seek to quantify how researchers theorize sexuality through a survey of 55 papers that specifically analyze sexuality bias in NLP systems. Following \citet{devinney_theories_2022}'s work on theories of gender in NLP, we document how sexuality is theorized ("\textit{what is sexuality?}") and operationalized ("\textit{what data represents sexuality?}"). We also consider if surveyed articles employ intersectional methodologies ("\textit{is sexuality bias measured simultaneously with other bias axes?}"). We find that a majority of the papers surveyed do not make their conception of sexuality clear, causing many papers to, among other things, conflate sexuality and gender, oversimplify sexuality, and adopt methodologies that position heterosexuality as the default. 

We conclude with recommendations for researchers who wish to measure sexuality-based bias in NLP systems. We argue that practitioners should make their working definition of sexuality explicit, and such a definition should consider intersectional identities when possible. Further, great care should be taken to construct datasets that draw on meaningful power dynamics rooted in real queer experiences. Finally, we take inspiration from queer theory to suggest possible paths forward that center queerness not just as a social demographic, but as an ideological framework for which to construct equitable NLP systems.

\section{Background}

We first outline related meta analyses of NLP bias research, then move to explore various theories of sexuality, drawing from community definitions and relevant theory. We also note how discussions of gender and intersectionality are necessary even if the focus of this analysis is sexuality. 

\subsection{Related Work}
\label{sec:related-work}
\citet{devinney_theories_2022} surveyed how NLP papers explicitly theorized and implicitly operationalized "gender" in bias research. They found that a majority of the papers surveyed "do not make their theorization of gender explicit" and that many papers "conflate sex characteristics, social gender, and linguistic gender" so as to exclude trans, non-binary, intersex, and intersectional gender identities. While their focus was on gender bias, we share similar motivations in investigating how queer people are under/misrepresented in NLP bias analyses.


Many studies have shown that certain bias measurement techniques rely on problematic assumptions that can undermine their effectiveness. \citet{blodgett_language_2020} found that motivations expressed by NLP bias articles tend to lack clear "normative reasoning," and that these papers' quantitative measures for evaluating bias often do not align with their intentions. \citet{blodgett_stereotyping_2021} further outlined how unclear articulations of social stereotypes lead to ambiguous and ineffective datasets for measuring bias in NLP tasks such as language modeling and coreference resolution. Additionally, \citet{gallegos_bias_2024} conducted an extensive survey of LLM bias evaluation and mitigation techniques, synthesizing potential limitations of both bias evaluation datasets and the methods/data structures used to extract bias. 

\subsection{Sexuality, Gender, and Social Theory}

A single, stable definition of sexuality is inherently illusive. From biology to sociology, theology to sexology, feminist studies to queer theory, there are vast differences between how sexuality is researched and theorized across disciplines. The very notion of sexuality as an identity characteristic, rather than a behavioral tendency, didn't take root in popular culture until the early 20th century \citep{katz_invention_2007}. As such, for the purposes of this study it would be improper to narrowly confine "sexuality" under a single all-encompassing definition. Therefore, we briefly outline how sexuality can be conceptualized along various continua, how gender and race are inextricably and historically linked to modern ideas of sexuality, and how queer and feminist theories may further trouble a stable definition of sexuality. 

\paragraph{Sexuality Spectra and Identities.} A useful heuristic for understanding modern sexuality is to place an individual's sexual and romantic preferences (or non-preferences) on various continua. One of the earliest attempts to categorize sexuality in this way was  "The Kinsey Scale," which saw sexuality as a variable mixture of homosexual and heterosexual urges \citep{kinsey_sexual_1948}. Moving slightly past finite identity categories, The Kinsey Scale allowed medical experts to categorize sexuality beyond a binary (homosexuality/heterosexuality) or trinary (homo/hetero/bisexuality) model. Similar methods of placing sexuality on a scale or spectrum are still popular today. Someone who feels a strong alignment with a certain sexual or romantic \textit{identity} may choose to self-identify their \textit{sexual/romantic orientation}, with common identities falling under broader homosexual/romantic, heterosexual/romantic, or bisexual/romantic categories. Asexuality or aromanticism are identities that describe less or no sexual/romantic attraction, and many see the "aro/ace" spectrum as another component of any person's sexuality. A somewhat comprehensive list of commonly used identities can be found on websites like the community-curated LGBTQIA+ Wiki\footnote{https://lgbtqia.fandom.com/wiki/}.




\paragraph{Gender.}
While gender is not the explicit focus of this paper, gender and sexuality are inextricably linked. At a minimum, many sexual identity labels are contingent upon a person's gender identity, meaning a discussion of sexuality-based NLP bias is incomplete without first analyzing how authors conceive and operationalize gender in NLP systems. As mentioned previously, \citet{devinney_theories_2022} investigated the ways in which NLP  researchers often employ binary notions of gender in bias research. Further, they find that many papers use "essentialist" frameworks that explicitly exclude trans and non-binary identities by constructing gender as immutable. 

Both binary and essentialist notions of gender can lead to improper conceptions of sexuality; a researcher aiming to study sexuality bias in NLP systems while failing to reconcile with gender presentations beyond static "man" or "woman" labels can lead to a lack of representation for gender-expansive sexualities. Further, the term "queer," as well as some Indigenous and non-Western identities (e.g., \textit{Two-Spirit} in Indigenous North American cultures or \textit{Hijra} in South-Asian cultures), further blur the line between gender and sexuality, a phenomenon that often complicates single-axis bias research. The ways researchers reconcile with the interplays between gender and sexuality are discussed in Section~\ref{sec:dis-conflation}.

\paragraph{Intersectionality.}
An \textit{intersectional} framework is essential for fully capturing how individuals with multiple marginalized identities experience oppression. "Intersectionality" was coined by \citet{crenshaw_demarginalizing_1989}, who articulates how a "single-axis framework" of identity does not fully account for the compounded way in which "Black women are subordinated." In the context of evaluating sexuality-based bias in NLP systems, this may mean considering the ways a technology may be more  biased against, say, "Black lesbian" identities versus simply "lesbian" identities. This is discussed further in Section~\ref{sec:dis-intersectionality}.

\paragraph{Queer and Feminist Theories.} 
To examine diverse conceptions of sexuality, we provide a brief overview of key theoretical frameworks from Feminist and Queer Studies. Namely, poststructuralist thought questions the stability of gender and sexual categories by theorizing identity as a socially constructed product of power relations. Famously, \citet{butler_gender_1990} articulates their concept of “gender performativity,” or the idea that gender is constructed through discursive practices and everyday critiques and relations. Further, Butler argues that sex, gender, and sexuality are cultural processes that are informed by heteronormativity\footnote{The theory that heterosexual culture is synonymous with normal ways of living \citep{warner_introduction_1991} or that heterosexuality is often assumed and mandated as the default sexual orientation \cite{rich_compulsory_1980}.}. Together with Foucault's \textit{The History of Sexuality} \citeyearpar{foucault_history_1978}, a critical historical response to the formation of sexuality categories, we can begin to understand sexual identities as institutionally constructed and socially reinforced phenomena. While these authors are suspicious of common identity categories, it is important to note that many theorists, particularly some Black feminist scholars, support theories of identity that emphasize one's relation to marginalization and enable coalitional empowerment \citep{ahmed_living_2017, cohen_punks_1997}. A discussion about the extent to which surveyed articles engage with sexuality theory is discussed throughout Section~\ref{sec:discussion} and suggestions for incorporating interdisciplinary theory in NLP work are presented in Section~\ref{sec:recommendations-future-work}.












\section{Survey}

\subsection{Method}
We collected papers concerned with sexuality-based bias in the field of Natural Language Processing by conducting multiple searches across two primary databases. In total, we obtained 55 papers between February and March of 2025 using the methods outlined by \citet{devinney_theories_2022} and \citet{blodgett_language_2020}.

First, we searched for papers with keywords relating to sexuality\footnote{See a more detailed explanation of how we search for sexuality in Appendix~\ref{app:search-terms}} together with "NLP"/"Natural Language Processing" and "bias" across the Association for Computational Machinery (ACM) Digital Library \footnote{https://dl.acm.org/}. Out of 71 search results, only five specifically dealt with sexuality bias in NLP systems. Next, we conducted a similar search across the Association for Computational Linguistics (ACL) Anthology \footnote{https://aclanthology.org/}, collecting 60 additional articles that fit our initial inclusion criteria. Of these initial 65 papers, ten were later excluded as they either did not specifically address bias within NLP systems or sufficiently address sexuality (as opposed to gender or other categories of bias). The complete list of all 55 papers are included in Appendix~\ref{app:survey-bib}. 

We then carefully read each paper in order to explore how NLP researchers conceptualize and operationalize sexuality in bias research. We categorize each paper by considering myriad primary factors, with each category summarized in Table~\ref{tab:schema}.

\begin{table*}
    \centering
    \begin{tabular}{l|l}
         \textbf{Category}& Description\\
         \hline
         \textbf{Sexuality Theory}& How is sexuality is theorized?\\
         \textbf{Sexuality Proxy}& What data is used to represent sexuality?\\
         \textbf{Sexuality Bias}& How is sexuality bias theorized or measured?\\
         \textbf{Sexuality Focus}& Is measuring sexuality bias the primary focus of the article?\\
         \textbf{Beyond Duality}& Does the article go beyond a queer/not queer binary comparison structure?\\
         \textbf{Intersectionality}& Is sexuality bias measured \textit{together} with other oppressions?\\
         \textbf{Language}& What language(s) are investigated?\\
         \textbf{Technology}& What technology is examined?\\
    \end{tabular}
    \caption{Categorization schema for surveyed papers.}
    \label{tab:schema}
\end{table*}
We also took supplemental notes for each paper, and document additional findings and trends in Section~\ref{sec:discussion}. The results from the survey are detailed below.

\subsection{Results}

\paragraph{Sexuality Theory.}We find a sizable variation in how sexuality is theorized across papers, evidenced in Table~\ref{tab:sexuality-theory}. Only 14 (25.5\%) articles are tagged as neither \textit{underspecified} or \textit{undefined}, meaning sexuality or queerness is explicitly defined to some extent. Three (5.4\%) papers are marked as \textit{undefined}, meaning sexuality is never defined and no definition can be reasonably assumed. A majority of the papers (38, or 69.1\%), however, are tagged as \textit{underspecified}. These papers do not explicitly define sexuality or queerness, but an implied framework can be inferred from methods or data. We find that about half the papers surveyed (n=27) consider more than three sexuality identities. However, only six (10.9\%) specifically acknowledge that sexuality is a spectrum or mention inherent difficulty in quantifying sexuality\footnote{Of these six papers, none attempt to operationalize complex definitions of sexuality beyond employing identity words for bias detection. Potential future work in this area is described in Section~\ref{rec:queer}.}.

\paragraph{Sexuality Proxy.}We find that papers often utilize one of two sexuality proxies (Table~\ref{tab:sexuality-proxy}): \textit{human annotation} (14, or 25.5\%) and \textit{identity word lists} (40, or 72.7\%).  Its important to note that only proxies explicitly stated by paper authors are counted in the survey. For example, papers that utilize human annotation of generated/collected text often contain more complex proxies for sexuality than simple word list (thereby necessitating manual annotation), but these proxies are not tabulated. Excluding proxies that involve annotation, a large majority of papers identify a set list of sexuality identity words (e.g., \textit{homosexual}, \textit{heterosexual}, \textit{asexual}, etc.) that are then used to conduct bias analyses (e.g., evaluating word embedding associations of straight vs. queer identity words; or analyzing the output of an LLM prompted with straight vs. queer identity words). One paper \citep{felkner_winoqueer_2023} uses  social media posts and online news articles about queer topics, and one paper \cite{stewart_whose_2024} explicitly uses \textit{grammatical gender}, \textit{pronoun}, and \textit{title} relations as a proxy for sexuality. 

\paragraph{Sexuality Bias.}Table~\ref{tab:sexuality-bias} details the large swath of bias measures employed across surveyed papers. Roughly a third (20, or 36.4\%) of articles utilize one or more automatic subjective language classifiers (tagged as either \textit{regard}, \textit{sentiment}, or \textit{toxicity/hate}) to compare outputs from various generative models. Other notable bias detection techniques observed were model \textit{performance} comparisons (15, or 27.3\%), generation \textit{likelihood} comparisons (14, or 25.5\%), \textit{counterfactual} data structures (13, or 23.6\%),   and \textit{word embedding} vector analyses (7, or 12.7\%). 

\begin{figure}
    \centering
    \includegraphics[width=1\linewidth]{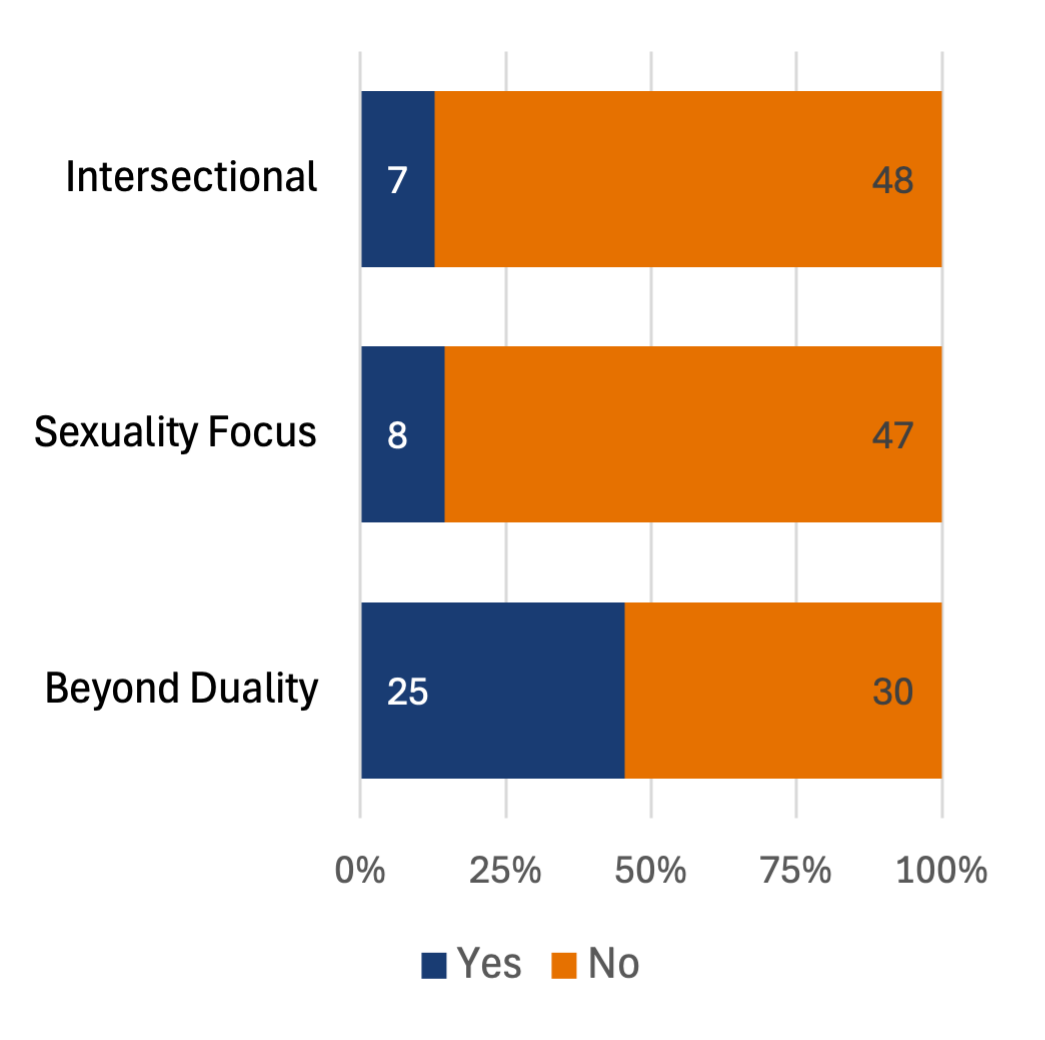}
    \caption{Additional paper classifications. \textbf{Intersectional}: Is sexuality bias measured together with other oppressions? The paper must consider multiple oppressed identities simultaneously. \textbf{Sexuality Focus}: Is measuring sexuality-based bias in NLP systems the primary focus of the article? \textbf{Beyond Duality}: Does the article go beyond a simple heterosexual/LGBTQ+ binary comparison structure?}
    \label{fig:moreclassifications}
\end{figure}

 Finally, Figure~\ref{fig:moreclassifications} reports the \textbf{Sexuality Focus}, \textbf{Beyond Duality}, and \textbf{Intersectionality} classifications. We find that a majority of papers do not have a sexuality focus or utilize an intersectional framework, meaning most papers consider multiple demographics but do so independently in their bias analysis. Additionally, we find that roughly half of the papers (25, or 45.5\%), go beyond a simple heterosexual/LGBTQ+ binary comparison structure. This means an article may consider multiple queer identities (and breaks down results by identity) or simply does not consider heterosexuality (common when only considering "marginalized" or "protected" groups).



\begin{table*}
  \centering
  \begin{tabular}{l|l||r}
    \textbf{Sexuality Theory} & Inclusion Criteria & \# \\
    \hline
        \textit{underspecified}&	does not explicitly/clearly define sexuality&	38
\\
        \textit{many identities}&	considers >3 sexuality identities&	27
\\
        \textit{aro/ace}&	considers identities along the aromantic/asexual spectrum&	16
\\
        \textit{binary}&	considers only homosexuality and heterosexuality&	6
\\
        \textit{spectrum}&	acknowledges sexuality is a spectrum/complex&	6
\\
        \textit{homosexuality only}&	considers only homosexuality&	4
\\
        \textit{trinary}&	considers only homosexuality, heterosexuality, and bisexuality&	3
\\
        \textit{romantic}&	considers romantic attraction as distinct from sexual attraction&	3
\\
        \textit{undefined}&	no clear framework&	3
\\
        \textit{culture-dependent}&	analyzes sexuality identity definitions across cultural/language contexts&	2
\\
        \textit{monolithic}&	sexuality only characterized by the word "LGBTQ+"&	1
\\
        \textit{sexuality is gender}&	sexuality identities suspiciously placed under gender category&	1\\
    
  \end{tabular}
  \caption{How is sexuality theorized across papers? Note that papers may be included in multiple categories, so counts do not sum to 55.}
  \label{tab:sexuality-theory}
\end{table*}

\begin{table*}
  \centering
  \begin{tabular}{l|l||r}
    \textbf{Sexuality Proxy} & Inclusion Criteria & \# \\
    \hline
        \textit{identity word list}&	uses a set list of identities  &	40
\\
 \textit{annotation (human)}& uses manual human annotation of generated/collected text&14
\\
        \textit{affiliation}&	uses collected text from social media spaces and news sources&	1
\\
        \textit{annotation (LLM)}&	uses automatic LLM annotation of generated/collected text&	1
\\
        \textit{grammatical gender rel.}&	uses gendered word relations&	1
\\
        \textit{pronoun rel.}&	uses pronoun relations &	1
\\
        \textit{titles}&	uses relationship titles &	1\\
    
  \end{tabular}
  \caption{What data is representative of sexuality? Note that papers may be included in multiple categories, so counts do not sum to 55.}
  \label{tab:sexuality-proxy}
\end{table*}

\begin{table*}
  \centering
  \begin{tabular}{l|l||r}
    \textbf{Sexuality Bias} & Inclusion Criteria & \# \\
    \hline
        \textit{toxicity/hate}&	score comparisons: uses an automatic toxicity/hate speech classifier&	16
\\
\textit{likelihood}&	compares the probability that a sentence/word was generated&	14
\\
\textit{counterfactual}&	compares stereotypical/non-stereotypical sentences&	13
\\
\textit{performance}&	evaluates a tool's correctness, considers >2 sexuality identities&	9
\\
\textit{word embeddings}&	compares sexuality identity word vectors&	7
\\
\textit{performance (binary)}&	evaluates a tool's correctness, considers hetero/homosexual identities&	6
\\
\textit{QA}&	asks: does a QA model prefer one answer choice over another?&	6
\\
\textit{occupation}&	uses occupation titles as a measure of bias&	5
\\
\textit{regard}&	score comparisons: uses "regard" \citep{sheng_woman_2019}&	5
\\
\textit{sentiment}&	score comparisons: uses an automatic sentiment classifier&	5
\\
\textit{associations}&	tests differences between identity category associations&	4
\\
\textit{data imbalance}&	considers imbalance in training data for certain identities&	2
\\
\textit{multiple}&	explicitly considers many dimensions of bias&	2
\\
\textit{translation}&	measures machine translation accuracy&	2
\\
\textit{allocational}&	concerned with differences in allocated resources&	1
\\
\textit{example}&	bias shown via provided examples&	1
\\
\textit{harm}&	compares number of harmful generations via manual annotation&	1
\\
\textit{heteronormative}&	addresses heteronormativity detection&	1\\
    
  \end{tabular}
  \caption{How is sexuality bias theorized and/or measured? Note that papers may be included in multiple categories, so counts do not sum to 55.}
  \label{tab:sexuality-bias}
\end{table*}







\section{Discussion}
\label{sec:discussion}

We find that an overwhelming number of papers surveyed do not make their theorization of sexuality explicit, indicating a reliance on assumed or normative conceptions of sexual/romantic practices and identities. Many of our findings are consistent with \citet{devinney_theories_2022}'s study of gender bias research, namely that authors often don't engage with gender/sexuality literature even in research that centers anti-queer bias. At a minimum, this under-citation of work outside of NLP overlooks essential interdisciplinary perspectives -- but at its worst, it can lead to incorrect conceptions of queerness that may perpetuate the very harm researchers are trying to mitigate. Below we outline some common  trends we observed while conducting the survey. 

\subsection{Gender/Sexuality Conflation}
\label{sec:dis-conflation}


Gender and sexuality are often used interchangeably in examined papers. Sometimes, collapsing queer gender and sexual identities is not done in a necessarily harmful manner. For example, a researcher who wishes to evaluate an NLP system's potential bias against queer people may test various identities, including both sexual and gender identities, and report results under a general "LGBTQ" bias category. This method of testing and reporting is broad and may not tell the reader any identity-specific information, but valid conclusions about bias may still be made. Often, however, sexuality and gender are conflated in ways that demonstrate under-informed or even misinformed theorizations of queerness.

The most basic type of gender/sexuality conflation is when an article classifies sexual identities as gender identities or gender identities as sexual identities. \citet{mukherjee_global_2023}, for example, does the latter when they characterize the type of bias "LGBTQ+" people face simply as "sexuality bias," grouping transgender people into specifically sexuality-based bias evaluations. \citet{hassan_unpacking_2021}, on the other hand, categorizes the words "...lesbian, gay, bisexual, [and] asexual..." under the category of "Gender Identity." 

Both of the authors above only explicitly consider either "sexuality" or "gender" bias in their methodology. In other words, sexuality and gender are not considered two separate constructs, leading to their misrepresentation. \citet{navigli_biases_2023} exemplifies a different, more troubling trend seen in some papers when the authors use "non-binary" as an example of a sexual orientation \textit{while also considering gender bias} elsewhere in the paper. Similarly, \citet{elsafoury_thesis_2023} classifies "transgender" under a "Sexual-orientation" table row even though a "Gender" category exists just two lines above. These papers showcase how researchers often attempt to maintain gender as the "binary case," even when other demographic factors are operationalized with multiple categories  \citep{devinney_theories_2022}. In other words, despite how both gender and sexuality are equally considered, trans and non-binary identities defy cisnormative, essentialist binaries and thus don't fit into gendered word lists that contain words like \textit{woman}, \textit{female}, \textit{sister}, or \textit{husband}. 

The word \textit{queer} reveals a particular struggle among NLP bias researchers. "Queer" can be used as an epithet (discussed in Section~\ref{sec:dis-protected-hate}), but can also refer to someone who identifies as a sexual \textit{or} gender minority. "Queer" is inherently anti-categorical and fluid, a phenomenon in tension with NLP systems that require immutable and mutually exclusive categories. As \citet{shah_i_2023} points out, queer identities are often marked as "dirty data" and discarded in AI systems because of this incompatibility, seen in one surveyed paper dealing with various demographic biases: 
\begin{quote}
\textit{"We exclude the descriptor ``queer'', an outlier because it falls in both the gender/sex and sexual orientation axes."}

\hfill -- \citet{costa-jussa_multilingual_2023}
\end{quote}
Recommendations for how queerness, both as an identity category or a poststructuralist framework, can be better conceptualized by NLP/AI researchers is discussed in Section~\ref{rec:queer}.


\subsection{Sexuality: Simplified}
\label{sec:dis-simplified}

Roughly a fifth (10, or 23.6\%) of the papers surveyed explicitly or implicitly exclude sexualities other than homosexuality or heterosexuality. Often, this exclusion of other identities is not addressed. Articles that do address the existence of more identities often do so only to then justify why they must be excluded because of some constraint. Similar to \citet{devinney_theories_2022}'s observation that binary gender is seen as a limitation, we find that binary sexuality is often listed as a technical limitation of bias analysis techniques. In one paper, the authors explain that their methodology only enables binary group comparisons while also acknowledging that queerness cannot be captured with a single word: 


\begin{quote}
\textit{"...we define pairs of theory specific words for each type of discrimination...Man-Woman, German-Foreigners, Straight-Gay, for sexist, xenophobic and homophobic prejudice respectively. Both gender and sexuality are spectra. We did not analyze here biases related to the rest social groups for simplicity reasons, as the methodology deals with group dualities."}

\hfill -- \citet{papakyriakopoulos_bias_2020}
\end{quote}
Another author frames a sexuality binary as a methodological strength: 

\begin{quote}
\textit{"[W]e examine the groups female and male for gender, Black and White for race, and gay and straight for sexual orientation. To constrain the scope of our analysis, we limit each demographic type to two classes, which, while unrepresentative of the real-world diversity, allows us to focus on more depth in analysis."}

\hfill -- \citet{sheng_woman_2019}
\end{quote}
Collapsing all non-straight sexualities under the "gay" umbrella demonstrates an oversimplification of queerness that deeply limits a study's capacity to measure complex bias. 





\subsection{Heteronormativity}
\label{sec:dis:heteronormativity}
We find that a significant portion of the articles surveyed employ heteronormative methodologies that position heterosexuality as the default/norm and queer sexualities as the "other." Heteronormative methodologies are not the same as intentionally examining queer discrimination or hate speech, which is discussed in Section~\ref{sec:dis-protected-hate}. Instead, we find that many papers employ techniques for extracting bias that privilege heterosexuality or are otherwise misaligned with queer ways of being.

\paragraph{Unmarkedness.} Linguists use the term "markedness" to describe how marginalized identities are highly visible in common language. Unmarked groups, on the other hand, are culturally privileged identities (such as \textit{white}, \textit{heterosexual}, or \textit{man}) that "gain a special, default status" because they are normatively assumed when an identity is not specified \citep{bucholtz_language_2005}. We find that heterosexuality is often unmarked or assumed as the default in more than one paper surveyed. For example, \citet{jiang_not_2020} tests how various identities are scored with a toxicity classifier, but does not include heterosexuality in their analysis despite considering  homosexual identities, as well as both marginalized and privileged racial and gender identities like black/white and man/woman. Although this omission privileges heterosexuality as the assumed norm, \citet{blodgett_stereotyping_2021} points out that simply adding “straight” words may not be an adequate solution. Since heterosexuality is often unmarked in natural language, using an automatic classifier to test the sentence "\textit{I am a straight man}" for toxicity, for example, may produce unreliable results, since \textit{straight man} is an artificially constructed example that likely does not naturally appear often in the classifier’s training data \cite{blodgett_stereotyping_2021}. 




\paragraph{Counterfactuals.} Counterfactual sentence structures are commonly used to test NLP systems for bias by perturbing identity words in a sentence while leaving the rest of the semantic meaning unchanged \cite{gallegos_bias_2024}. A drastic change in a model output in response to the changed identity word could indicate bias. For example, CrowS-Pairs \cite{nangia_crows-pairs_2020}, a dataset from one of the papers surveyed, uses the following to probe for sexuality-based model bias:
\begin{quote}
\textit{"Most \underline{gay} / \underline{straight} men are not manly so they do not have thick beards."}

\hfill -- \citet{nangia_crows-pairs_2020}
\end{quote}
Sentences like this suffer from the unmarkedness problem described above, but they also structurally frame heterosexuality and queerness as distinct opposites. While the authors' goal is to examine model bias with respect to stereotypical/anti-stereotypical sentences, doing so in this way places heterosexuality as the touchstone sexuality for which queerness should be compared to. \citet{blodgett_stereotyping_2021} point out various faults with CrowS-Pairs, such as how the dataset often presents illogical or problematic stereotype pairs. One example we found places queerness at odds with whiteness: 
\begin{quote}
\textit{"Then they disobeyed god by being \underline{lesbians} / \underline{white} and beginning to kiss."}

\hfill -- \citet{nangia_crows-pairs_2020}
\end{quote}
\citet{nangia_crows-pairs_2020}  frames \textit{white} as an appropriate "contrasting advantaged group" to \textit{lesbians}, a "historically disadvantaged group." Beyond the issue that \textit{white} is a racial identity that should not be used to test for sexuality bias, the assertion that white, rather than straight, is the most logical contrasting word to lesbian demonstrates that heterosexuality is thoroughly ingrained as the default in US culture, so much so that whiteness and heterosexuality are implicitly theorized as overlapping or mutually reinforcing categories of privilege.


Counterfactuals may have the potential to go beyond centering heterosexuality, but no paper that uses this structure attempts to do so. \citet{felkner_winoqueer_2023} employ contrasting pairs and include myriad queer identities, but, similar to \citet{nangia_crows-pairs_2020}, always use heterosexuality as the default with which to compare. We believe researchers can uncover more nuanced model bias by decentering heterosexuality and moving beyond direct comparison of pre-defined identity keywords, which is discussed more in Secion~\ref{rec:queer}.





\subsection{Intersectionality}
\label{sec:dis-intersectionality}
Using \citet{crenshaw_mapping_1991}'s terminology, "intragroup differences" are often ignored or conflated under monolithic identities in a majority of the papers surveyed. Specifically, 87.3\% (n=48) of papers do not adopt an intersectional framework of analysis, meaning that sexuality bias is often measured independently of, say, racial bias. Most papers do not acknowledge how bias may be compounded for identities that face multifaceted societal oppression. Some authors acknowledge the existence of intersectionality, but exclude it from their methodology because of technical limitations or simplifications. For example: 
\begin{quote}
\textit{"Examples demonstrating intersectional examples are valuable...but we find that allowing multiple tag choices dramatically lowers the reliability of the tags."}

\hfill -- \citet{nangia_crows-pairs_2020}
\end{quote}
Other authors may reference intersectional theory but fail to meaningfully engage with intersectional identities in their analysis, often treating multiple identities separately rather than examining how they interact \citep{hassan_unpacking_2021, mukherjee_global_2023}. Finally, we observe that articles that study word embeddings are often limited in their intersectional analysis, simply because token-based embeddings can usually only capture single words. Thus, multifaceted identities that require multiple words to describe may not be within the scope of a word-embedding analysis. We echo \citet{gallegos_bias_2024} in questioning the effectiveness of word embedding–based bias analysis for evaluating real-world discrimination.


\subsection{Hate Contexts}
\label{sec:dis-protected-hate}
Sexuality is sometimes theorized only through the lens of hate or as an attribute that contributes to hurtful language. For example, at least three papers surveyed \citep{elsafoury_comparative_2022, nozza_honest_2021, nozza_measuring_2022} use the Hurtlex lexicon \citep{cabrio_hurtlex_2018}, which frames homosexuality as a component of hate speech.  Papers that use benchmarks such as this one may classify text containing the word "gay" as hurtful, even if the tone of the text is neutral or benign. 

Reclaimed words, or words that have historically functioned as epithets that have been reworked/embraced by marginalized groups, also cause trouble for NLP researchers. Oftentimes, these words are either exclusively seen as signs of hate speech or simply excluded from analysis because of their multiple meanings. Both of these implementations are problematic for words like \textit{queer}, which, while historically used as a slur, is commonly used by people in the LGBTQ+ community. Even the word \textit{gay} may trigger negative associations in bias analyses because of historical pejorative uses of the word \citep{dhingra_queer_2023}. By failing to account for the reclamation and evolving meanings of these words, NLP researchers risk perpetuating outdated biases and misrepresenting the lived experiences of marginalized communities, ultimately preventing more nuanced analysis. 

\subsection{Cultural Contexts and Western Hegemony}

We found that nearly all of the papers surveyed examine sexuality through a Western, often US-centric lens, unless the paper specifically focuses on non-Western cultural contexts. Most papers do not acknowledge their Western focus, but those that do often list it simply as a limitation. For example: 

\begin{quote}
\textit{"Our study is also very Western-centric: we study only English models/datasets and test for biases considered normatively pressing in Western research."}

\hfill -- \citet{steed_upstream_2022}
\end{quote}
The two papers that were marked as employing \textit{culture-dependent} theories of sexuality \citep{fort_your_2024, gamboa_filipino_2024}, however, specifically examine how Western benchmark datasets may be both linguistically and culturally translated into different global contexts. These studies attempt to transform English/Western stereotypes about sexual minorities into stereotypes that can be used for multilingual bias testing. 




\section{Recommendations and Future Work}
\label{sec:recommendations-future-work}

As we have demonstrated, current efforts to quantify sexuality bias in NLP systems often reproduce limited theories of sexuality that fail to account for the complexity of queer identities. Simultaneously, heterosexuality is often positioned as the default or baseline with which to measure all other relationship structures. To address this, we provide a list of recommendations for the NLP community based on our observations throughout the study, as well as point to future research that may bridge the gap between NLP and queer theory.  

\subsection{Make Theory of Sexuality Explicit}
\label{rec:theory-explicit}

We adapt \citet{devinney_theories_2022}'s call for researchers to clearly state their theoretical frameworks around gender when addressing gender-related NLP bias. Sexuality, like gender, should not be taken for granted and assumed as an obvious or self-explanatory category. While it may be impossible (and unnecessary) to fully capture all relevant academic and community theories of sexuality, we don't believe it is unreasonable to make clear how sexuality is operationalized and measured in work that specifically addresses sexuality-based bias. This is in line with \citet{blodgett_language_2020}'s assertion that language technology researchers should clearly state why certain system biases are harmful, in what ways, and to \textit{whom}. For example, simply providing a list of identity words is insufficient without an explanation of why certain identities are included over others.

\subsection{Allow for Intersectionality and Nuance}
\label{rec:nuance}

In the same vein as clearly explaining how sexuality is theorized, researchers should leave space for more complex identities and nuanced methodologies. In some cases, grouping all queer identities under a broad "LGBTQ+" bias category may be adequate, especially if sexuality is one of many dimensions being considered, but many papers surveyed do so even if their methodologies allow for unambiguous, identity-specific reporting. 

When explaining results, we encourage researchers to break down nuanced signs of bias and situate these results in the context of how the bias was measured. When this isn't possible, researchers should refrain from grouping all queer sexualities together when reporting results, especially if more specific identities were actually used to extract model bias. In short, the more explicit a study can be in describing exactly who is harmed by a biased system, the better. 

We also encourage researchers to adopt intersectional methodologies. Oftentimes, researchers who create template-based prompts to measure model bias only measure a single demographic axis at a time. On the other hand, researchers who annotate real-world data, where social categories are often entangled, may still collapse text into a single category of bias. Both of these strategies ignore the ways queer Black women, for example, experience unique and compounded discrimination \citep{crenshaw_demarginalizing_1989}. Researchers often defend the use of single-dimension frameworks by arguing that they allow for unambiguous analysis (see Section~\ref{sec:dis-intersectionality}). However, we argue that NLP practitioners should reframe their methodologies to account for the specific and distinct biases faced by individuals at the intersections of multiple forms of oppression, rather than framing these identities as indeterminate and analytically imprecise.

\subsection{Construct Meaningful Power Dynamics}
\label{rec:power}

We emphasize \citet{gallegos_bias_2024}'s recommendation that researchers "construct metrics to reflect real world power dynamics" when evaluating NLP technologies for bias. Specifically, datasets and methodologies that measure anti-queer bias should be rooted in the actual experiences of people within the queer community. This necessarily requires significant engagement with both relevant interdisciplinary literature and diverse members of the queer community that face different levels of homophobia, racism, mysogony, and transphobia. 

In a similar vein, large language models should not be used as a substitute for human annotators. As \citet{felkner_gpt_2024} found, GPT-based dataset annotation fails to capture harmful stereotypes that are specific to the queer community. By abstracting away queer (human) involvement in dataset annotation,  meaningful and nuanced power dynamics are missed, severely limiting the effectiveness of the bias detection technology.

\subsection{Decenter Heterosexuality as the Default}
\label{rec:hetero}

As discussed in Section~\ref{sec:dis:heteronormativity}, many papers surveyed use methodologies that center heterosexuality as the default. In fact,  heterosexuality is often not operationalized as a sexual identity, evidenced by the handful of papers surveyed that claim to analyze sexuality-based bias but only consider queer sexualities. While having a clear theory of sexuality, as discussed above, may help address this issue, we observe that heteronormativity can still permeate a paper’s methodology. For instance, WinoQueer \citep{felkner_winoqueer_2023} encouragingly makes clear their theorization of sexuality and reports detailed results broken down by identity category. However, their methodology is based off CrowS-Pairs \citep{nangia_crows-pairs_2020}, which employ counterfactual stereotypical/anti-stereotypical pairs to evaluate bias. While WinoQueer is able to show clear model bias by comparing myriad queer identities to "a corresponding non-LGBTQ+ identity," this approach is constrained by its assumption that queer identities in fact have "corresponding," mutually exclusive opposites. We ask, then, what nuanced biases could be uncovered if we don't take heterosexuality as the baseline for which to compare? We discuss potential strategies for this in the following section.



\subsection{Explore Queer Methodologies}
\label{rec:queer}
\begin{quote}
\textit{"...a decidedly queer approach can question the very logics of visibility with which algorithmic systems and AI are trained."} 

\hfill -- \citet{klipphahn-karge_queer_2024}
\end{quote}
Recently, queer theorists and other interdisciplinary scholars have argued that queerness, both as a loose grouping of sexual and gender identities and a framework with which to understand anti-categorical, fluid conceptions of life and power, is somewhat at odds with current AI implementations. As the above quote demonstrates, scholars such as \citet{klipphahn-karge_queer_2024} argue that proposed solutions for debiasing these systems, such as increasing data representation or considering additional identities, do not do enough to tackle systematic marginalization. AI systems, they emphasize, heavily rely on strict, defined, immutable, and mutually exclusive categories -- all of which are at odds with queer realities. How, then, could we work towards an anti-heteronormative AI future that centers queer existence?

One possible strategy for NLP researchers is to deliberately reduce dependence on fixed or stable identity categories. A significant majority of papers surveyed (40, or 72.7\%) employ identity word lists as a proxy for sexuality/queerness. While using identity words to extract model bias leaves little room for contextual ambiguity, constructing a supposedly 'comprehensive' identity list inevitably excludes less common or non-categorizable sexualities from analysis. Further, \citet{butler_gender_1990} and \citet{foucault_history_1978} assert that gender and sexual categories are rooted in (and reinforced by) social structures of power that work to preserve heteronormativity. Identity categories should not be completely discarded, but these observations indicate that there may be room for a more queer approach to NLP bias measurement. 

Along these lines, \citet{shah_i_2023} has suggested that AI researchers work to limit their reliance on highly prevalent categories in their infrastructures. Doing so, Shah argues, would work towards "queering the node," thus centering the collective in AI networks.  Encouragingly, researchers such as \citet{attanasio_entropy-based_2022} have pointed to methods of this nature that "do not require any a-priori knowledge of identity terms," relying instead on more automatic methods to detect and mitigate unwanted bias. Inspired by this, we encourage future work that simultaneously centers the lived experiences of queer folk while challenging NLP's reliance on stable identity categories as the basis for bias evaluation.











\section{Conclusion}
\begin{quote}
\textit{"...if analysis could build a queer utopia alone, we would not still be here."} 

\hfill -- \citet{keyes_inconclusion_2023}
\end{quote}
Through a survey of 55 papers that concern sexuality bias in Natural Language Processing, we found that a majority of NLP researchers do not clearly articulate what sexuality is or why particular proxies for sexuality should be used. Further, we found that few papers operationalize nuanced or intersectional methodologies that account for the specific realities of those who may face the greatest systematic bias from NLP systems. We proposed multiple actionable recommendations, distilled into the following checklist for NLP bias researchers:

\begin{itemize}
    \item[\ding{51}] Make theory of sexuality explicit, including why certain identities are included/excluded.
    \item[\ding{51}] Refrain from treating queerness as a monolith; adopt intersectional methodologies and make precise statements about exactly who is harmed by a biased system.
    \item[\ding{51}] Engage with relevant interdisciplinary literature, community members, and stakeholders throughout the research process. 
\end{itemize}

Finally, we encourage sociotechnical research that aims to "queer" NLP methodologies by challenging fundamental assumptions about identity and bias.



\section*{Acknowledgments}

I am deeply grateful to my advisor, Professor Briana Morrison, for her invaluable guidance and unwavering support in helping bring this project to fruition. I am also indebted to Professor Isabel Gonzales, a close mentor who inspired this work and taught me to seek transformative solutions in pursuit of a queer future. 



\bibliography{references}


\appendix
\section{Sexuality Search Terms}
\label{app:search-terms}

Below are the terms we used to search for articles that relate to sexuality-based NLP bias. We acknowledge that these terms do not encompass all dimensions of sexuality, but believe they are sufficient enough to search for articles in the field that address sexuality/queer bias.  
\subsection{ACL Anthology}
\label{app:ACL}
Because of limitations in the ACL Anthology search feature, namely that only 100 articles are accessible per unique search, we conducted two searches for the ACL Anthology database. The first search consisted of general "sexuality" words, while the second specifically searched for "queer" words. Splitting the search in this way allowed us to find numerous articles that only deal with either sexuality or queerness.
\begin{itemize}
    \item \textit{}{Sexuality (general) keywords:} [\texttt{sexual orientation}, \texttt{sexuality}]
    \item \textit{}{Queer identity keywords:} [\texttt{queer}, \texttt{lgb}, \texttt{lgbt}, \texttt{lgbtq}, \texttt{lgbtqia+}]
\end{itemize}

\subsection{ACM Digital Library}
The ACM Digital Library allows for more complex search queries (such as using * as a multi-character wildcard), simplifying our search for queer identities.
\begin{itemize}
    \item \textit{}{Sexuality/queer identity keywords:} [\texttt{sexual orientation}, \texttt{sexuality}, \texttt{queer}, \texttt{lgb*}]
\end{itemize}

\section{Additional Survey Results}
\label{app:addSurveyResults}
Additional information collected about all surveyed papers, including what language the paper investigated and what technology was examined, is available in Table~\ref{tab:language} and \ref{tab:technology}. 

\begin{table}
  \centering
  \begin{tabular}{l|r}
    \textbf{Language} & \# \\
    \hline
        \textit{English (only)}	&		42	\\
        \textit{single language (non-English)}	&		6	\\
        \textit{>1 language} &	7	\\
    
  \end{tabular}
  \caption{What language(s) are investigated in each paper?}
  \label{tab:language}
\end{table}

\begin{table}
  \centering
  \begin{tabular}{l|r}
    \textbf{Technology} & \# \\
    \hline
        \textit{Convolutional Neural Networks} &	2	\\
        \textit{Large Language Models} &	44	\\
        \textit{Machine Translation} &	3	\\
        \textit{Support Vector Machines} &	2	\\
        \textit{Word/Sentence Embeddings} &	8	\\
        \textit{other} &	6	\\
    
  \end{tabular}
  \caption{What technology does the paper examine for bias?}
  \label{tab:technology}
\end{table}

\section{Full Bibliography of Surveyed Papers}
\label{app:survey-bib}

\begin{hangparas}{.11in}{1}

{\small 

Soumya Barikeri, Anne Lauscher, Ivan Vulić, and Goran Glavaš. 2021. RedditBias: A Real-World Resource for Bias Evaluation and Debiasing of Conversational Language Models. In \textit{Proceedings of the 59th Annual Meeting of the Association for Computational Linguistics and the 11th International Joint Conference on Natural Language Processing (Volume 1: Long Papers)}, pages 1941–1955, Online. Association for Computational Linguistics.  \\

Lisa Bauer, Karthik Gopalakrishnan, Spandana Gella, Yang Liu, Mohit Bansal, and Dilek Hakkani-Tur. 2022. Analyzing the Limits of Self-Supervision in Handling Bias in Language. In \textit{Findings of the Association for Computational Linguistics: EMNLP 2022}, pages 7372–7386, Abu Dhabi, United Arab Emirates. Association for Computational Linguistics. \\

Selma Bergstrand and Björn Gambäck. 2024. Detecting and Mitigating LGBTQIA+ Bias in Large Norwegian Language Models. In \textit{Proceedings of the 5th Workshop on Gender Bias in Natural Language Processing (GeBNLP)}, pages 351–364, Bangkok, Thailand. Association for Computational Linguistics. \\

Marta Costa-jussà, Pierre Andrews, Eric Smith, Prangthip Hansanti, Christophe Ropers, Elahe Kalbassi, Cynthia Gao, Daniel Licht, and Carleigh Wood. 2023. Multilingual Holistic Bias: Extending Descriptors and Patterns to Unveil Demographic Biases in Languages at Scale. In \textit{Proceedings of the 2023 Conference on Empirical Methods in Natural Language Processing}, pages 14141–14156, Singapore. Association for Computational Linguistics. \\

Paula Czarnowska, Yogarshi Vyas, and Kashif Shah. 2021. Quantifying Social Biases in NLP: A Generalization and Empirical Comparison of Extrinsic Fairness Metrics. \textit{Transactions of the Association for Computational Linguistics}, 9:1249–1267. \\

Aida Mostafazadeh Davani, Mohammad Atari, Brendan Kennedy, and Morteza Dehghani. 2023. Hate Speech Classifiers Learn Normative Social Stereotypes. \textit{Transactions of the Association for Computational Linguistics}, 11:300–319. \\

Fatma Elsafoury. 2022. Darkness can not drive out darkness: Investigating Bias in Hate SpeechDetection Models. In \textit{Proceedings of the 60th Annual Meeting of the Association for Computational Linguistics: Student Research Workshop}, pages 31–43, Dublin, Ireland. Association for Computational Linguistics. \\

Fatma Elsafoury. 2023. Thesis Distillation: Investigating The Impact of Bias in NLP Models on Hate Speech Detection. In \textit{Proceedings of the Big Picture Workshop}, pages 53–65, Singapore. Association for Computational Linguistics. \\

Fatma Elsafoury, Steve R. Wilson, Stamos Katsigiannis, and Naeem Ramzan. 2022a. SOS: Systematic Offensive Stereotyping Bias in Word Embeddings. In \textit{Proceedings of the 29th International Conference on Computational Linguistics}, pages 1263–1274, Gyeongju, Republic of Korea. International Committee on Computational Linguistics. \\

Fatma Elsafoury, Steven R. Wilson, and Naeem Ramzan. 2022b. A Comparative Study on Word Embeddings and Social NLP Tasks. In \textit{Proceedings of the Tenth International Workshop on Natural Language Processing for Social Media}, pages 55–64, Seattle, Washington. Association for Computational Linguistics. \\

David Esiobu, Xiaoqing Tan, Saghar Hosseini, Megan Ung, Yuchen Zhang, Jude Fernandes, Jane Dwivedi-Yu, Eleonora Presani, Adina Williams, and Eric Michael Smith. 2023. ROBBIE: Robust Bias Evaluation of Large Generative Language Models. \\

Virginia Felkner, Ho-Chun Herbert Chang, Eugene Jang, and Jonathan May. 2023. WinoQueer: A Community-in-the-Loop Benchmark for Anti-LGBTQ+ Bias in Large Language Models. In \textit{Proceedings of the 61st Annual Meeting of the Association for Computational Linguistics (Volume 1: Long Papers)}, pages 9126–9140, Toronto, Canada. Association for Computational Linguistics. \\

Shangbin Feng, Chan Young Park, Yuhan Liu, and Yulia Tsvetkov. 2023. From Pretraining Data to Language Models to Downstream Tasks: Tracking the Trails of Political Biases Leading to Unfair NLP Models. In \textit{Proceedings of the 61st Annual Meeting of the Association for Computational Linguistics (Volume 1: Long Papers)}, pages 11737–11762, Toronto, Canada. Association for Computational Linguistics. \\

Karen Fort, Laura Alonso Alemany, Luciana Benotti, Julien Bezançon, Claudia Borg, Marthese Borg, Yongjian Chen, Fanny Ducel, Yoann Dupont, Guido Ivetta, Zhijian Li, Margot Mieskes, Marco Naguib, Yuyan Qian, Matteo Radaelli, Wolfgang S. Schmeisser-Nieto, Emma Raimundo Schulz, Thiziri Saci, Sarah Saidi, et al. 2024. Your Stereotypical Mileage May Vary: Practical Challenges of Evaluating Biases in Multiple Languages and Cultural Contexts. In \textit{Proceedings of the 2024 Joint International Conference on Computational Linguistics, Language Resources and Evaluation (LREC-COLING 2024)}, pages 17764–17769, Torino, Italia. ELRA and ICCL. \\

Lance Calvin Lim Gamboa and Mark Lee. 2024. Filipino Benchmarks for Measuring Sexist and Homophobic Bias in Multilingual Language Models from Southeast Asia. arXiv:2412.07303 [cs]. \\

Salvatore Greco, Ke Zhou, Licia Capra, Tania Cerquitelli, and Daniele Quercia. 2024. NLPGuard: A Framework for Mitigating the Use of Protected Attributes by NLP Classifiers. \textit{Proceedings of the ACM on Human-Computer Interaction}, 8(CSCW2):1–25. \\

Saad Hassan, Matt Huenerfauth, and Cecilia Ovesdotter Alm. 2021. Unpacking the Interdependent Systems of Discrimination: Ableist Bias in NLP Systems through an Intersectional Lens. In \textit{Findings of the Association for Computational Linguistics: EMNLP 2021}, pages 3116–3123, Punta Cana, Dominican Republic. Association for Computational Linguistics. \\

Carolin Holtermann, Anne Lauscher, and Simone Ponzetto. 2022. Fair and Argumentative Language Modeling for Computational Argumentation. In \textit{Proceedings of the 60th Annual Meeting of the Association for Computational Linguistics (Volume 1: Long Papers)}, pages 7841–7861, Dublin, Ireland. Association for Computational Linguistics. \\

Jiachen Jiang and Soroush Vosoughi. 2020. Not Judging a User by Their Cover: Understanding Harm in Multi-Modal Processing within Social Media Research. In \textit{Proceedings of the 2nd International Workshop on Fairness, Accountability, Transparency and Ethics in Multimedia}, pages 6–12, Seattle WA USA. ACM. \\

Abhishek Kumar, Sarfaroz Yunusov, and Ali Emami. 2024. Subtle Biases Need Subtler Measures: Dual Metrics for Evaluating Representative and Affinity Bias in Large Language Models. In \textit{Proceedings of the 62nd Annual Meeting of the Association for Computational Linguistics (Volume 1: Long Papers)}, pages 375–392, Bangkok, Thailand. Association for Computational Linguistics. \\

Mingyu Ma, Jiun-Yu Kao, Arpit Gupta, Yu-Hsiang Lin, Wenbo Zhao, Tagyoung Chung, Wei Wang, Kai-Wei Chang, and Nanyun Peng. 2024. Mitigating Bias for Question Answering Models by Tracking Bias Influence. In \textit{Proceedings of the 2024 Conference of the North American Chapter of the Association for Computational Linguistics: Human Language Technologies (Volume 1: Long Papers)}, pages 4592–4610, Mexico City, Mexico. Association for Computational Linguistics. \\

Mamta Mamta, Rishikant Chigrupaatii, and Asif Ekbal. 2024. BiasWipe: Mitigating Unintended Bias in Text Classifiers through Model Interpretability. In \textit{Proceedings of the 2024 Conference on Empirical Methods in Natural Language Processing}, pages 21059–21070, Miami, Florida, USA. Association for Computational Linguistics. \\

Marta Marchiori Manerba and Sara Tonelli. 2021. Fine-Grained Fairness Analysis of Abusive Language Detection Systems with CheckList. In \textit{Proceedings of the 5th Workshop on Online Abuse and Harms (WOAH 2021)}, pages 81–91, Online. Association for Computational Linguistics. \\

Anjishnu Mukherjee, Chahat Raj, Ziwei Zhu, and Antonios Anastasopoulos. 2023. Global Voices, Local Biases: Socio-Cultural Prejudices across Languages. In \textit{Proceedings of the 2023 Conference on Empirical Methods in Natural Language Processing}, pages 15828–15845, Singapore. Association for Computational Linguistics. \\

Jimin Mun, Emily Allaway, Akhila Yerukola, Laura Vianna, Sarah-Jane Leslie, and Maarten Sap. 2023. Beyond Denouncing Hate: Strategies for Countering Implied Biases and Stereotypes in Language. In \textit{Findings of the Association for Computational Linguistics: EMNLP 2023}, pages 9759–9777, Singapore. Association for Computational Linguistics. \\

Nikita Nangia, Clara Vania, Rasika Bhalerao, and Samuel R. Bowman. 2020. CrowS-Pairs: A Challenge Dataset for Measuring Social Biases in Masked Language Models. In \textit{Proceedings of the 2020 Conference on Empirical Methods in Natural Language Processing (EMNLP)}, pages 1953–1967, Online. Association for Computational Linguistics. \\

Roberto Navigli, Simone Conia, and Björn Ross. 2023. Biases in Large Language Models: Origins, Inventory, and Discussion. \textit{Journal of Data and Information Quality}, 15(2):1–21. \\

Isar Nejadgholi, Esma Balkir, Kathleen Fraser, and Svetlana Kiritchenko. 2022. Towards Procedural Fairness: Uncovering Biases in How a Toxic Language Classifier Uses Sentiment Information. In \textit{Proceedings of the Fifth BlackboxNLP Workshop on Analyzing and Interpreting Neural Networks for NLP}, pages 225–237, Abu Dhabi, United Arab Emirates (Hybrid). Association for Computational Linguistics. \\

Shangrui Nie, Michael Fromm, Charles Welch, Rebekka Görge, Akbar Karimi, Joan Plepi, Nazia Mowmita, Nicolas Flores-Herr, Mehdi Ali, and Lucie Flek. 2024. Do Multilingual Large Language Models Mitigate Stereotype Bias? In \textit{Proceedings of the 2nd Workshop on Cross-Cultural Considerations in NLP}, pages 65–83, Bangkok, Thailand. Association for Computational Linguistics. \\

Debora Nozza, Federico Bianchi, and Dirk Hovy. 2021. HONEST: Measuring Hurtful Sentence Completion in Language Models. In \textit{Proceedings of the 2021 Conference of the North American Chapter of the Association for Computational Linguistics: Human Language Technologies}, pages 2398–2406, Online. Association for Computational Linguistics. \\

Debora Nozza, Federico Bianchi, Anne Lauscher, and Dirk Hovy. 2022. Measuring Harmful Sentence Completion in Language Models for LGBTQIA+ Individuals. In \textit{Proceedings of the Second Workshop on Language Technology for Equality, Diversity and Inclusion}, pages 26–34, Dublin, Ireland. Association for Computational Linguistics. \\

Swetasudha Panda, Ari Kobren, Michael Wick, and Qinlan Shen. 2022. Don`t Just Clean It, Proxy Clean It: Mitigating Bias by Proxy in Pre-Trained Models. In \textit{Findings of the Association for Computational Linguistics: EMNLP 2022}, pages 5073–5085, Abu Dhabi, United Arab Emirates. Association for Computational Linguistics. \\

Orestis Papakyriakopoulos, Simon Hegelich, Juan Carlos Medina Serrano, and Fabienne Marco. 2020. Bias in word embeddings. In \textit{Proceedings of the 2020 Conference on Fairness, Accountability, and Transparency}, pages 446–457, Barcelona Spain. ACM. \\

Alicia Parrish, Angelica Chen, Nikita Nangia, Vishakh Padmakumar, Jason Phang, Jana Thompson, Phu Mon Htut, and Samuel Bowman. 2022. BBQ: A hand-built bias benchmark for question answering. In \textit{Findings of the Association for Computational Linguistics: ACL 2022}, pages 2086–2105, Dublin, Ireland. Association for Computational Linguistics. \\

Chahat Raj, Anjishnu Mukherjee, Aylin Caliskan, Antonios Anastasopoulos, and Ziwei Zhu. 2024. BiasDora: Exploring Hidden Biased Associations in Vision-Language Models. In \textit{Findings of the Association for Computational Linguistics: EMNLP 2024}, pages 10439–10455, Miami, Florida, USA. Association for Computational Linguistics. \\

Shaina Raza, Ananya Raval, and Veronica Chatrath. 2024. MBIAS: Mitigating Bias in Large Language Models While Retaining Context. In \textit{Proceedings of the 14th Workshop on Computational Approaches to Subjectivity, Sentiment, \& Social Media Analysis}, pages 97–111, Bangkok, Thailand. Association for Computational Linguistics. \\

Timo Schick, Sahana Udupa, and Hinrich Schütze. 2021. Self-Diagnosis and Self-Debiasing: A Proposal for Reducing Corpus-Based Bias in NLP. \textit{Transactions of the Association for Computational Linguistics}, 9:1408–1424. \\

Sagi Shaier, Kevin Bennett, Lawrence Hunter, and Katharina Kann. 2023. Emerging Challenges in Personalized Medicine: Assessing Demographic Effects on Biomedical Question Answering Systems. In \textit{Proceedings of the 13th International Joint Conference on Natural Language Processing and the 3rd Conference of the Asia-Pacific Chapter of the Association for Computational Linguistics (Volume 1: Long Papers)}, pages 540–550, Nusa Dua, Bali. Association for Computational Linguistics. \\

Emily Sheng, Kai-Wei Chang, Prem Natarajan, and Nanyun Peng. 2020. Towards Controllable Biases in Language Generation. In \textit{Findings of the Association for Computational Linguistics: EMNLP 2020}, pages 3239–3254, Online. Association for Computational Linguistics. \\

Emily Sheng, Kai-Wei Chang, Prem Natarajan, and Nanyun Peng. 2021. Societal Biases in Language Generation: Progress and Challenges. In \textit{Proceedings of the 59th Annual Meeting of the Association for Computational Linguistics and the 11th International Joint Conference on Natural Language Processing (Volume 1: Long Papers)}, pages 4275–4293, Online. Association for Computational Linguistics. \\

Emily Sheng, Kai-Wei Chang, Premkumar Natarajan, and Nanyun Peng. 2019. The Woman Worked as a Babysitter: On Biases in Language Generation. In \textit{Proceedings of the 2019 Conference on Empirical Methods in Natural Language Processing and the 9th International Joint Conference on Natural Language Processing (EMNLP-IJCNLP)}, pages 3407–3412, Hong Kong, China. Association for Computational Linguistics. \\

Jisu Shin, Hoyun Song, Huije Lee, Soyeong Jeong, and Jong Park. 2024. Ask LLMs Directly, “What shapes your bias?”: Measuring Social Bias in Large Language Models. In \textit{Findings of the Association for Computational Linguistics: ACL 2024}, pages 16122–16143, Bangkok, Thailand. Association for Computational Linguistics. \\

Eric Michael Smith, Melissa Hall, Melanie Kambadur, Eleonora Presani, and Adina Williams. 2022. “I`m sorry to hear that”: Finding New Biases in Language Models with a Holistic Descriptor Dataset. In \textit{Proceedings of the 2022 Conference on Empirical Methods in Natural Language Processing}, pages 9180–9211, Abu Dhabi, United Arab Emirates. Association for Computational Linguistics. \\

Ryan Steed, Swetasudha Panda, Ari Kobren, and Michael Wick. 2022. Upstream Mitigation Is \textit{N}ot All You Need: Testing the Bias Transfer Hypothesis in Pre-Trained Language Models. In \textit{Proceedings of the 60th Annual Meeting of the Association for Computational Linguistics (Volume 1: Long Papers)}, pages 3524–3542, Dublin, Ireland. Association for Computational Linguistics. \\

Ian Stewart and Rada Mihalcea. 2024. Whose wife is it anyway? Assessing bias against same-gender relationships in machine translation. In \textit{Proceedings of the 5th Workshop on Gender Bias in Natural Language Processing (GeBNLP)}, pages 365–375, Bangkok, Thailand. Association for Computational Linguistics. \\

Asahi Ushio, Yi Zhou, and Jose Camacho-Collados. 2023. Efficient Multilingual Language Model Compression through Vocabulary Trimming. In \textit{Findings of the Association for Computational Linguistics: EMNLP 2023}, pages 14725–14739, Singapore. Association for Computational Linguistics. \\

Juan Vásquez, Scott Andersen, Gemma Bel-enguix, Helena Gómez-adorno, and Sergio-luis Ojeda-trueba. 2023. HOMO-MEX: A Mexican Spanish Annotated Corpus for LGBT+phobia Detection on Twitter. In \textit{The 7th Workshop on Online Abuse and Harms (WOAH)}, pages 202–214, Toronto, Canada. Association for Computational Linguistics. \\

Juan Vásquez, Gemma Bel-Enguix, Scott Andersen, and Sergio-Luis Ojeda-Trueba. 2022. HeteroCorpus: A Corpus for Heteronormative Language Detection. In \textit{Proceedings of the 4th Workshop on Gender Bias in Natural Language Processing (GeBNLP)}, pages 225–234, Seattle, Washington. Association for Computational Linguistics. \\

Sean Xie, Saeed Hassanpour, and Soroush Vosoughi. 2024. Addressing Healthcare-related Racial and LGBTQ+ Biases in Pretrained Language Models. In \textit{Findings of the Association for Computational Linguistics: NAACL 2024}, pages 4451–4464, Mexico City, Mexico. Association for Computational Linguistics. \\

Abdelrahman Zayed, Goncalo Mordido, Ioana Baldini, and Sarath Chandar. 2024. Why Don`t Prompt-Based Fairness Metrics Correlate? In \textit{Proceedings of the 62nd Annual Meeting of the Association for Computational Linguistics (Volume 1: Long Papers)}, pages 9002–9019, Bangkok, Thailand. Association for Computational Linguistics. \\

Guanhua Zhang, Bing Bai, Junqi Zhang, Kun Bai, Conghui Zhu, and Tiejun Zhao. 2020. Demographics Should Not Be the Reason of Toxicity: Mitigating Discrimination in Text Classifications with Instance Weighting. In \textit{Proceedings of the 58th Annual Meeting of the Association for Computational Linguistics}, pages 4134–4145, Online. Association for Computational Linguistics. \\

Jiaxu Zhao, Meng Fang, Zijing Shi, Yitong Li, Ling Chen, and Mykola Pechenizkiy. 2023. CHBias: Bias Evaluation and Mitigation of Chinese Conversational Language Models. In \textit{Proceedings of the 61st Annual Meeting of the Association for Computational Linguistics (Volume 1: Long Papers)}, pages 13538–13556, Toronto, Canada. Association for Computational Linguistics. \\

Jiaxu Zhao, Zijing Shi, Yitong Li, Yulong Pei, Ling Chen, Meng Fang, and Mykola Pechenizkiy. 2024. More than Minorities and Majorities: Understanding Multilateral Bias in Language Generation. In \textit{Findings of the Association for Computational Linguistics: ACL 2024}, pages 9987–10001, Bangkok, Thailand. Association for Computational Linguistics.  \\

Yi Zhou, Danushka Bollegala, and Jose Camacho-Collados. 2024. Evaluating Short-Term Temporal Fluctuations of Social Biases in Social Media Data and Masked Language Models. In \textit{Proceedings of the 2024 Conference on Empirical Methods in Natural Language Processing}, pages 19693–19708, Miami, Florida, USA. Association for Computational Linguistics. \\

Muitze Zulaika and Xabier Saralegi. 2025. BasqBBQ: A QA Benchmark for Assessing Social Biases in LLMs for Basque, a Low-Resource Language. In \textit{Proceedings of the 31st International Conference on Computational Linguistics}, pages 4753–4767, Abu Dhabi, UAE. Association for Computational Linguistics. \\

}
\end{hangparas}



\end{document}